\documentclass[12pt]{article}
\usepackage{amssymb}
\usepackage{amsmath}
\usepackage{amsfonts}

\newcommand{\R}{\mathbb{R}}
\newcommand{\C}{\mathbb{C}}
\newcommand{\Z}{\mathbb{Z}}
\newcommand{\N}{\mathbb{N}}

\newcommand{\be}{\begin{equation}}
\newcommand{\ee}{\end{equation}}
\newcommand{\bea}{\begin{eqnarray}}
\newcommand{\eea}{\end{eqnarray}}
\newcommand{\nn}{\nonumber}
\newcommand{\kt}{\rangle}
\newcommand{\br}{\langle}

\newcommand{\ed}{\end{document}}
\newcommand{\bbr}{\br\!\br}
\newcommand{\kkt}{\kt\!\kt}

\newcommand{\QQ}{{\cal Q}}
\begin{document}

\title{Statistical Origin of Pseudo-Hermitian Supersymmetry
and Pseudo-Hermitian Fermions}

\author{\\
Ali Mostafazadeh
%\thanks{E-mail address: amostafazadeh@ku.edu.tr}\\
\\ Department of Mathematics, Ko\c{c} University,\\
Rumelifeneri Yolu, 34450 Sariyer,\\
Istanbul, Turkey\\ \\
amostafazadeh@ku.edu.tr}
\date{ }
\maketitle

\begin{abstract}
We show that the metric operator for a pseudo-supersymmetric
Hamiltonian that has at least one negative real eigenvalue is
necessarily indefinite. We introduce pseudo-Hermitian fermion
(phermion) and abnormal phermion algebras and provide a pair of
basic realizations of the algebra of $N=2$ pseudo-supersymmetric
quantum mechanics in which pseudo-supersymmetry is identified with
either a boson-phermion or a boson-abnormal-phermion exchange
symmetry. We further establish the physical equivalence
(non-equivalence) of phermions (abnormal phermions) with ordinary
fermions, describe the underlying Lie algebras, and study
multi-particle systems of abnormal phermions. The latter provides
a certain bosonization of multi-fermion systems.
\end{abstract}
%\vspace{2mm}
%\begin{center}
%~~~~~~~~PACS numbers: 11.30.-j, 11.40.-q, 03.65.Pm, 98.80.Qc
%\end{center}
%\vspace{2mm}

%\baselineskip=24pt

%\textheight = 22cm \topskip = -1cm \topmargin = -1cm
%\newpage

\section{Introduction}

Supersymmetry entered theoretical physics as the symmetry allowing
for the exchange or mixing of the fermionic and bosonic degrees of
freedom in certain quantum field theories \cite{susy-qft}. The
subsequent attempts \cite{witten-81-82} to understand the issue of
the spontaneous supersymmetry-breaking led to the discovery of
supersymmetric quantum mechanics (SQM)
\cite{susy-review,susy-review2}. Since its inception in the early
1980s, SQM has become a focus of attention for its various
physical applications \cite{susy-spin,susy-review2,susy-book} and
mathematical implications \cite{susy-math}.

The impact of SQM in theoretical physics and mathematics motivated
the introduction and investigation of various generalizations of
SQM. Most of these generalizations, including parasupersymmetric
\cite{para-susy}, orthosupersymmetric \cite{ortho-susy,jpa},
$q$-deformed \cite{q-susy} and fractional \cite{f-susy}
supersymmetric quantum mechanics, are algebraic in nature in the
sense that they are defined in terms of an underlying operator
algebra that generalizes that of SQM. The main guiding principle
in generalizing the algebra of SQM has been to replace the role
played by the fermionic degree of freedom in SQM with that of a
parafermionic, orthofermionic, or $q$-fermionic degree of
freedom.\footnote{The exchange symmetry of an order 2
parafermion-paraboson pair turns out to be a centrally extended
$N=4$ supersymmetry \cite{ijmpa-96-3}. The $q$-boson-fermion and
$q$-boson-$q$-fermion exchange symmetries have also been
considered in the literature \cite{qb-f,qb-qf}.}

In addition to these algebraic generalizations, there is also a
class of topological generalizations of supersymmetry called
topological symmetries \cite{top-susy1} -- \cite{top-susy4}. The
latter are defined in terms of certain conditions on the spectral
degeneracy structure of the Hamiltonian so that the corresponding
system mimics the topological properties of the supersymmetric
systems. Specifically, to each quantum system possessing a
topological symmetry, there is associated a set of topological
invariants that generalize the Witten index of SQM. Topological
symmetries also have underlying operator algebras, and as it is
shown in \cite{top-susy2,top-susy3} the operator algebras of most
of the above-mentioned algebraic generalizations of SQM may be
recovered as those of the topological symmetries.

Recently, the present author has introduced yet a third class of
generalizations of supersymmetry called pseudo-Hermitian
supersymmetry or pseudo-supersymmetry \cite{p1,p4}. The latter may
be viewed as a geometric generalization of supersymmetry, because
its definition relies on allowing for the adjoint of its
generator(s) to be defined in terms of a possibly indefinite inner
product (hence the metric) on the Hilbert space. The main
practical motivation for the introduction of pseudo-supersymmetry
is that it provides a general framework \cite{p4} that encompasses
all the attempts at generating non-Hermitian Hamiltonians with a
real spectrum by intertwining Hermitian Hamiltonians \cite{old}.

The algebra of pseudo-SQM of order $2N$ is given by
    \be
    \QQ_a^2=0,~~~~~~~[\QQ_a,H]=0,~~~~~~~~
    \{\QQ_a,\QQ_b^\sharp\}=2\delta_{ab}H,
    \label{ph-susy}
    \ee
where $\QQ_a$ with $a\in\{1,2,\cdots,N\}$ are distinct generators
of pseudo-supersymmetry, $H$ is the Hamiltonian, and for any
linear operator $A$ acting in the Hilbert space (${\cal H}$) the
operator $A^\sharp$ stands for the pseudo-adjoint of $A$. The
latter is defined in terms of an invertible bounded Hermitian
(self-adjoint) linear operator $\eta:{\cal H}\to{\cal H}$
according to \cite{p1}
    \be
    A^\sharp:=\eta^{-1}A^\dagger\eta,
    \label{ph}
    \ee
where $A^\dagger$ stands for the ordinary adjoint of $A$, i.e.,
the unique operator satisfying $\br\cdot|A\:\cdot\kt= \br
A^\dagger\:\cdot|\cdot\kt$ with $\br\cdot|\cdot\kt$ denoting the
inner product of ${\cal H}$. Clearly, for $\eta=1$,
$A^\sharp=A^\dagger$ and the algebra~(\ref{ph-susy}) coincides
with that of SQM of order $2N$.

The operator $\eta$, which is a possibly indefinite
operator\footnote{An indefinite operator is a Hermitian operator
whose spectrum includes both strictly negative and strictly
positive real numbers.}, defines a possibly indefinite inner
product \cite{p1,indefinite} on ${\cal H}$, namely
$\bbr\cdot,\cdot\kkt_\eta:= \br\cdot|\eta\:\cdot\kt$. As a result,
it is sometimes referred to as a metric operator. The
pseudo-adjoint $A^\sharp$ is the adjoint of $A$ with respect to
the inner product $\bbr\cdot,\cdot\kkt_\eta$, i.e., $A^\sharp$ is
the unique operator satisfying $\bbr\cdot|A\:\cdot\kkt_\eta= \bbr
A^\sharp\cdot|\cdot\kkt_\eta$.

The main purpose of this article is to seek for the statistical
origin of pseudo-supersymmetry. In particular, we will consider a
Hamiltonian of the form\footnote{This is the simplest Hamiltonian
that allows for the study of the statistical origin of
supersymmetry and its various generalization, in particular
pseudo-supersymmetry. It is also motivated by field theoretical
considerations.}
    \be
    H=E(N+{\cal N})
    \label{H}
    \ee
where $E$ is a real constant, $N$ is the boson number operator,
and ${\cal N}$ is the number operator for a single degree of
freedom with an initially unknown algebra ${\cal A}$ of creation
and annihilation operators. We will then enforce the condition
that this system possesses an $N=2$ pseudo-supersymmetry. This
leads to two different sets of defining relations for ${\cal A}$
depending on whether the metric operator is definite or
indefinite.

The first and probably the most natural candidate for ${\cal A}$
is the pseudo-Hermitian generalization of the fermion algebra,
namely
    \be
    \alpha_+^2=\alpha_+^{\sharp 2}=0,~~~~~~
    \{\alpha_+,\alpha_+^\sharp\}=1,
    \label{normal-ph-fermion}
    \ee
where $\alpha_+^\sharp:=\eta^{-1}\alpha_+^\dagger\eta$ and
$\alpha_+$ are respectively the creation and annihilation
operators of what we call a {\em pseudo-Hermitian fermion} or
simply a {\em phermion}.\footnote{Note that pseudo-Hermitian
fermions are different from  the pseudo-fermions of
Ref.~\cite{beckers}. Similarly there is no direct relationship
between the notion of pseudo-Hermitian supersymmetry that we
abbreviate as pseudo-supersymmetry \cite{p1,p4} and the
boson-pseudo-fermion exchange symmetry that is also called
pseudo-supersymmetry \cite{beckers,qv}.} The corresponding number
operator is given by
    \be
    {\cal N}_+= \alpha_+^\sharp\alpha_+,
    \label{ph-fermion-number}
    \ee
so we set ${\cal N}={\cal N}_+$.

Again for $\eta=1$, (\ref{normal-ph-fermion}) reduces to the
fermion algebra. In general, as we will see, the algebra
(\ref{normal-ph-fermion}) does not support an indefinite metric
operator $\eta$. Furthermore, for any positive- or
negative-definite $\eta$, we can map (\ref{normal-ph-fermion})
onto the fermion algebra by a similarity transformation. Hence in
this case ${\cal N}$ is equivalent to the fermion number operator
and the pseudo-supersymmetry of $H$ is equivalent to ordinary
$N=2$ supersymmetry.

For an indefinite metric operator a simple and natural candidate
for the algebra ${\cal A}$ -- that is compatible with the
pseudo-supersymmetry of $H$ -- is the defining operator algebra of
what we propose to call an {\em abnormal pseudo-Hermitian fermion}
or simply an {\em abnormal phermion}. It is given by
    \be
    \alpha_-^2=\alpha_-^{\sharp 2}=0,~~~~~~
    \{\alpha_-,\alpha_-^\sharp\}=-1,
    \label{ph-fermion}
    \ee
where $\alpha_-^\sharp$ and $\alpha_-$ are respectively the
abnormal phermion creation and annihilation operators.

The above choice of the terminology has been partly adopted from a
paper of Sudarshan, namely \cite{sudarshan-61}, on
indefinite-metric quantum mechanics \cite{ph-indefinite,cjp1},
where he considers the following algebra of creation
($a_-^\dagger$) and annihilation ($a_-$) operators and refers to
it as the ``abnormal commutation relations''.
    \be
    [a_-,a_-^\dagger]=-1.
    \label{abnormal}
    \ee

Note that abnormal fermions --- whose defining algebra would
correspond to replacing $\sharp$ by $\dagger$ (setting $\eta=1$)
in (\ref{ph-fermion}) --- do not exist. This is because
$\{\alpha_-,\alpha_-^\dagger\}$ being a positive operator cannot
be equated to $-1$. It is the notion of the pseudo-adjoint
\cite{p1} that allows for considering a fermionic analog of
Sudarshan's abnormal bosonic degrees of freedom
\cite{sudarshan-61}.

The present article is organized as follows. In Section~2, we
present a general discussion of the algebras of creation and
annihilation operators and review the basic realization of $N=2$
SQM using the Hamiltonian~(\ref{H}) with ${\cal N}$ being the
fermion number operator. In section~3, we summarize the main
properties of $N=2$ pseudo-supersymmetric systems and establish a
previously unnoticed spectral consequence of pseudo-supersymmetry.
In Section~4, we describe the pseudo-supersymmetry of the
Hamiltonian~(\ref{H}) for the case that ${\cal N}$ is identified
with the phermion number operator and show that in this case
$\eta$ cannot be indefinite. Here we also demonstrate the physical
equivalence of phermions and fermions. In Section~5, we explore
the basic properties of abnormal phermions and discuss their role
in obtaining a fundamental realization of $N=2$
pseudo-supersymmetry with an indefinite metric operator. In
Section~6, we elucidate the group theoretical basis of the
phermion and abnormal phermion algebras. Finally in Section~7, we
offer a summary of our findings and present our concluding
remarks.

\section{Algebras of Creation and Annihilation Operators and the
Basic Realization of $N=2$ SQM}

Consider a complex $*$-algebra\footnote{A complex $*$-algebra is a
complex vector space ${\cal C}$ endowed with an associative
multiplication (which makes it into a complex associative algebra)
and a map $*:{\cal C}\to{\cal C}$ with the following properties.
Let $z\in\C$ (with complex-conjugate $\bar z$) and $a,b\in{\cal
C}$ be arbitrary and denote $*(a)$ by $a^*$. Then (1)~$(a^*)^*=a$,
so that $*$ is an involution; (2)~$(za)^*=\bar z a^*$ and
$(a+b)^*=a^*+b^*$, so that $*$ is antilinear; (3)~$(ab)^*=b^*a^*$,
\cite{fell-doran}.} generated by three elements: $c, c^*, n$ and
subject to the relations
    \be
    n^*=n,~~~~~~[c,n]=c.
    \label{a1}
    \ee
Then $c^*, c$, and $n$ are respectively called the creation,
annihilation, and number operators of a particle whose statistical
properties are determined by supplementing (\ref{a1}) with one or
more additional relations among the generators, namely
    \be
    P_\ell(c,c^*,n)=0,~~~~~~~~~~~\ell\in\{1,2,\cdots,r\},
    \label{a2}
    \ee
where $r\in\Z^+$ and $P_\ell:\C^3\to\C$ is a polynomial for each
$\ell$. Note that the relations obtained by applying $*$ to both
sides of (\ref{a1}) and (\ref{a2}) are understood to hold as well.
For example, in this way we obtain from (\ref{a1}) the relation
    \be
    [c^*,n]=-\,c^*.
    \label{a1-star}
    \ee

Let ${\cal C}$ denote the complex $*$-algebra generated by $c,
c^*, n$ and subject to the relations (\ref{a1}) and (\ref{a2}).
Suppose that  ${\cal C}$ admits faithful irreducible Hilbert space
representations $({\cal H},\rho)$ with Hilbert (representation)
space ${\cal H}$ and the representation map $\rho:{\cal C}\to {\rm
End}({\cal H})$, where ${\rm End}({\cal H})$ stands for the
complex associative algebra of linear operators acting in ${\cal
H}$. Furthermore suppose that there exists such a representation
of ${\cal C}$ that is also a $*$-representation \cite{fell-doran}
with respect to some possibly indefinite \cite{indefinite} inner
product $\bbr\cdot,\cdot\kkt$ on ${\cal H}$ that may differ from
its defining inner product $\br\cdot|\cdot\kt$. This means that
for all $v\in{\cal C}$, $\rho(v^*)=\rho(v)^\ddagger$, where
$\ddagger$ denotes the adjoint with respect to the inner product
$\bbr\cdot,\cdot\kkt$, i.e., for all $L\in{\rm End}({\cal H})$,
$L^\ddagger\in{\rm End}({\cal H})$ is the unique linear operator
defined by $\bbr\cdot,L\,\cdot\kkt=\bbr
L^\ddagger\cdot,\cdot\kkt$. Then ${\cal C}$ together with $({\cal
H},\rho)$ describe a physical particle\footnote{A physical
particle may or may not be a fundamental particle. The
qualification ``physical'' means that the ensuing mathematical
structure has the potential to describe physical systems
displaying effective particle-like behavior, e.g., quasi-particles
of condensed matter physics.}, and ${\cal C}$ is called the
abstract algebra of creation and annihilation operators of this
particle.

The typical examples of the above notion of a physical particle
are bosons and fermions. The abstract operator algebra ${\cal B}$
for a boson is determined by (\ref{a1}) and the relations
    \bea
    &&n=c^*c,
    \label{a4}\\
    &&[c,c^*]=1.
    \label{boson}
    \eea
Similarly, the abstract operator algebra ${\cal F}$ for a fermion
is determined by (\ref{a1}),  (\ref{a4}) and
    \bea
    &&c^2=0,
    \label{fermion-1}\\
    &&\{c,c^*\}=1.
    \label{fermion-2}
    \eea
It is not difficult to see that the above description of a
physical particle applies to parafermions and parabosons
\cite{para,greenberger-messiah}, orthofermins
\cite{ortho-susy,jpa}, $q$-deformed fermions \cite{qb-qf},
$q$-deformed bosons \cite{q-deformed-boson} and their
generalizations \cite{general}.

It is well-known that both ${\cal B}$ and ${\cal F}$ have (up to
equivalence) unique (unitary) $*$-irreducible
representations.\footnote{A unitary representation is a
$*$-representation with the inner product on the representation
space ${\cal H}$ identified with its defining positive-definite
inner product $\br\cdot|\cdot\kt$.} The unitary irreducible
representation of ${\cal B}$ is (up to similarity transformations)
given by ${\cal H}=L^2(\R)$ endowed with the usual $L^2$-inner
product and the representation map $\rho_b:{\cal B}\to{\rm
End}(L^2(\R))$ (uniquely) determined by $\rho_b(c)=(X+iP)/\sqrt
2=:a$ where $X$ and $P$ are normalized (dimensionless) position
and momentum operators satisfying $[X,P]=i1$. The fact that
$(L^2(\R),\rho_b)$ is a faithful unitary representation of ${\cal
B}$ implies that $\rho_b(1)=1$ and
$\rho_b(c^*)=\rho_b(c)^\dagger=a^\dagger$. In particular in this
representation, (\ref{boson}) takes the form
    \be
    [a,a^\dagger]=1,
    \label{boson-cr}
    \ee
and the number operator $n$ is represented by
    \be
    N:=a^\dagger a.
    \label{boson-number}
    \ee

The unique faithful unitary irreducible representation of the
fermion algebra ${\cal F}$ is two-dimensional. The representation
space is ${\cal H}=\C^2$ equipped with the Euclidean inner
product, and the representation map $\rho_f:{\cal F}\to{\rm
End}(\C^2)$ is (uniquely) determined by
    \be
    \rho_f(c):=\left(\begin{array}{cc}
            0 & 1 \\
            0 & 0\end{array}\right)=:\alpha.
    \label{f1}
    \ee
Again the fact that $(\C^2,\rho_f)$ is a faithful unitary
representation of ${\cal F}$ implies that $\rho_f(1)=1$ and
$\rho_f(c^*)=\rho_f(c)^\dagger=\alpha^\dagger$. Hence, in this
representation (\ref{fermion-1}) and (\ref{fermion-2})
respectively take the form
    \be
    \alpha^2=0,~~~~~~~~~~\{\alpha,\alpha^\dagger\}=1,
    \label{fermion-cr}
    \ee
and the fermion number operator $n$ becomes
    \be
    {\cal N}:=\alpha^\dagger\alpha.
    \label{fermion-number}
    \ee

Because of the uniqueness of the faithful unitary irreducible
representations of both ${\cal B}$ and ${\cal F}$, we will
suppress the representation maps $\rho_b$ and $\rho_f$ and use $a$
for $c$ in the case of a boson and $\alpha$ for $c$ in the case of
a fermion. This allows us to identify the boson and fermion
algebras with (\ref{boson-cr}) and (\ref{fermion-cr})
respectively.

Now, consider the boson-fermion oscillator \cite{supermanifolds}
whose Hamiltonian is given by (\ref{H}) with $E>0$, $N$ denoting
the boson number operator (\ref{boson-number}), and ${\cal N}$
labelling the fermion number operator (\ref{fermion-number}). The
Hilbert space ${\cal H}$ of this system is $L^2(\R)\otimes\C^2$
which is isomorphic as a Hilbert space to $L^2(\R)\oplus L^2(\R)$.

If we postulate the relative bose statistics
\cite{greenberger-messiah}:
    \be
    [a,\alpha]=[a^\dagger,\alpha]=0,
    \label{re-boson}
    \ee
the Hamiltonian (\ref{H}) of the boson-fermion oscillator commutes
with
    \be
    \tau:=1-2{\cal N}.
    \label{tau}
    \ee
Furthermore, in view of (\ref{fermion-cr}), (\ref{boson-number}),
(\ref{fermion-number}) and (\ref{re-boson}), we have $\tau^2=1$
and $\tau^\dagger=\tau$. Hence $\tau$ is a grading operator
splitting the Hilbert space into a direct sum of its eigenspaces:
    \be
    {\cal H}_\pm:=\{\psi\in{\cal H}|\tau\psi=\pm\psi\},
    \label{eg-space}
    \ee
i.e., ${\cal H}={\cal H}_+\oplus {\cal H}_-$. (The elements of)
${\cal H}_+$ and ${\cal H}_-$ are respectively called the bosonic
and fermionic (state vectors) Hilbert spaces.

The ground state of $H$ is represented by the unique state vector
$|0,+\kt$ eliminated by both $a$ and $\alpha$. In position
representation it takes the form
$\pi^{-1/4}e^{-x^2/2}\mbox{\scriptsize$\left(\begin{array}{c}1\\
0\end{array}\right)$}$. The vectors $|n,\epsilon\kt:= (n
!)^{-1/2}a^{\dagger n}(\alpha^{\dagger})^{(1-\epsilon)/2}
|0,+\kt$, with $n\in\N:=\{0,1,2,\cdots\}$, form an orthonormal
basis of ${\cal H}_\epsilon$ where $\epsilon\in\{-1,1\}$. The
spectrum of $H$ consists of a nondegenerate zero eigenvalue with
eigenvector $|0,+\kt$ and doubly degenerate positive eigenvalues,
$E_n=En$, with the pair of linearly independent eigenvectors
$(|n,+\kt,|n-1,-\kt)$, where $n\in\Z^+$.

It is well-known that this system has an $N=2$ supersymmetry
generated by
    \be
    \QQ=\sqrt{2E}~a^\dagger\alpha.
    \label{q-susy}
    \ee
Using (\ref{boson-cr}), (\ref{fermion-cr}), and (\ref{re-boson}),
we can easily check that
    \be
    \QQ^2=0,~~~~~~~[\QQ,H]=0,~~~~~~~~
    \{\QQ,\QQ^\dagger\}=2H.
    \label{susy}
    \ee
Note that here as well as in the rest of this article we use the
same symbol `$\dagger$' to denote the adjoint of operators acting
in $L^2(\R)$, $\C^2$, and ${\cal H}=L^2(\R)\otimes\C^2$.

As indicated by the expression (\ref{q-susy}), the physical
meaning of the above-mentioned supersymmetry of the boson-fermion
oscillator is the symmetry allowing for the exchange of bosonic
and fermionic states. This is conveniently summarized by the
identity $\{\QQ,\tau\}=0$.

\section{Consequences of Pseudo-Supersymmetry}

The defining ingredients of $N=2$ pseudo-SQM are the associated
operator algebra:
    \be
    \QQ^2=0,~~~~~~~[\QQ,H]=0,~~~~~~~~
    \{\QQ,\QQ^\sharp\}=2H,
    \label{N=2-ph-susy}
    \ee
and the $\Z_2$-graded structure of the Hilbert space ${\cal H}$.
The latter is specified through the existence of a grading
operator $\tau:{\cal H}\to{\cal H}$ that generalizes (\ref{tau})
in the sense that it satisfies $\tau^{-1}=\tau=\tau^\dagger$ and
hence leads to the decomposition ${\cal H}={\cal H}_+\oplus{\cal
H}_-$ of the Hilbert space where ${\cal H}_\pm$ are defined
according to (\ref{eg-space}). Furthermore, $\tau$ anticommutes
with the pseudo-supersymmetry generator $\QQ$ and commutes with
the metric operator $\eta$:
    \be
    \{\tau,\QQ\}=0=[\tau,\eta].
    \label{grading}
    \ee

These properties of the grading operator $\tau$ allow for a
canonical representation of the $N=2$ pseudo-supersymmetry in
which the restriction of $\QQ$ to ${\cal H}_-$ vanishes. This in
turn implies that $\QQ^\sharp$ has vanishing restriction onto
${\cal H}_+$. The situation is best described using the following
two-component representation of the Hilbert space in which a state
vector $|\psi\kt\in{\cal H}$ having $|\psi,\pm\kt$ definite
grading components, i.e., $|\psi\kt=|\psi,+\kt+|\psi,-\kt$ with
$|\psi,\pm\kt\in{\cal H}_\pm$, is represented as
$|\psi\kt=\mbox{\scriptsize{$
\left(\begin{array}{c}|\psi,+\kt\\|\psi,-\kt\end{array}\right)$}}$.
In this representation we have
    \be
    \QQ=\left(\begin{array}{cc}
    0 & 0\\
    D & 0\end{array}\right),~~~
    \eta=\left(\begin{array}{cc}
    \eta_+ & 0\\
    0 & \eta_-\end{array}\right),~~~
%    \label{2-comp-1}\\
    \QQ^\sharp=\left(\begin{array}{cc}
    0 & D^\sharp\\
    0 & 0\end{array}\right),~~~
    H=\left(\begin{array}{cc}
    H_+ & 0\\
    0 & H_-\end{array}\right),
    \label{2-comp-2}
    \ee
where $D:=\QQ|_{{\cal H}_+}$, $\eta_\pm:=\eta|_{{\cal H}_\pm}$,
$D^\sharp:=\QQ^\sharp|_{{\cal H}_-}=\eta_+^{-1}D^\dagger\eta_-$,
$H_+:=H|_{{\cal H}_+}=D^\sharp D/2$, and $H_-:=H|_{{\cal
H}_-}=D\,D^\sharp/2$.

Perhaps the most important feature of pseudo-supersymmetric
systems is that similarly to the ordinary supersymmetric systems
the nonzero eigenvalues are doubly degenerate \cite{p4}. However,
the spectrum need not be nonnegative.\footnote{Negative energy
eigenvalues also arise in some of the algebraic generalization of
supersymmetry such as parasupersymmetry \cite{para-susy} and
polynomial (nonlinear) supersymmetry \cite{non-lin-susy}. These
generalizations do not seem to be directly related to
pseudo-supersymmetry, for the latter does not restrict the
Hamiltonian to be Hermitian.} It may even include
complex-conjugate pairs of eigenvalues \cite{p4}. The argument for
the presence of double degeneracy for nonzero eigenvalues is
identical with the one for supersymmetry: Because $H$ and $\tau$
commute they may be simultaneously diagonalized; one may choose to
work with the eigenvectors of the Hamiltonian that have definite
grading. Now, suppose $|\psi_n,+\kt\in{\cal H}_+$ is such an
eigenvector with definite grading ($+$) and eigenvalue $E_n\neq
0$. Then $|\psi_n,+\kt$ pairs with the eigenvector $|\psi_n,-\kt:=
\QQ|\psi_n,+\kt\in{\cal H}_-$ which has the opposite grading ($-$)
and the same eigenvalue $E_n$.

The following is another simple consequence of
pseudo-supersymmetry that has, however, no supersymmetric analog.
    \begin{itemize}
\item[] {\bf Theorem:} Let $H$ be a diagonalizable
    pseudo-supersymmetric Hamiltonian with a discrete spectrum
    and $\eta$ be a metric operator defining the pseudo-adjoint
    $\QQ^\sharp$ of the pseudo-supersymmetry generator $\QQ$.
    If $H$ has a negative real eigenvalue, then $\eta$ is
    necessarily an indefinite operator.
\item[] {\bf Proof:} Suppose $H$ has nonzero real eigenvalues.
Because it is diagonalizable and commutes with $\tau$, there is a
complete basis of eigenvectors of $H$ with definite grading. Let
$|\psi_n,+\kt\in{\cal H}_+$ be such an eigenvector with a nonzero
real eigenvalue $E_n$, then as we showed above so is
$|\psi_n,-\kt:= \QQ|\psi_n,+\kt\in{\cal H}_-$. Now, compute
    \bea
    \bbr \psi_n,-|\psi_n,-\kkt_\eta&=&
    \br\psi_n,+|\QQ^\dagger\eta\QQ|\psi_n,+\kt\nn
    =\br\psi_n,+|\eta\QQ^\sharp\QQ|\psi_n,+\kt\nn\\
    &=&2E_n\br\psi_n,+|\eta|\psi_n,+\kt=
    2E_n\bbr \psi_n,+|\psi_n,+\kkt_\eta,
    \label{spect}
    \eea
where we have made use of (\ref{ph}) and (\ref{N=2-ph-susy}). As
shown in \cite{p1}, $|\psi_n,+\kt$ is $\eta$-orthogonal to all the
eigenvectors of $H$ having an eigenvalue different from $E_n$.
Furthermore,
$\bbr\psi_n,-|\psi_n,+\kkt_\eta=\br\psi_n,-|\eta|\psi_n,+\kt=
\br\psi_n,-|\eta_+|\psi_n,+\kt=0$. This is best seen using the
two-component representation~(\ref{2-comp-2}). In particular, it
implies that if $\bbr \psi_n,+|\psi_n,+\kkt_\eta=0$, then
$|\psi_n,+\kt$ belongs to the kernel of $\eta$. This contradicts
the fact that $\eta$ is an invertible operator. Hence, $\bbr
\psi_n,+|\psi_n,+\kkt_\eta\neq 0$. In view of this relation and
Eq.~(\ref{spect}), $\bbr \psi_n,+|\psi_n,+\kkt_\eta$ and $\bbr
\psi_n,-|\psi_n,-\kkt_\eta$ have the same sign if $E_n>0$. They
have the opposite sign if $E_n<0$. In particular, the presence of
a negative real eigenvalue implies that $\eta$ must be an
indefinite operator.~~~$\square$

\item[] {\bf Corollary:} Let $H$ be as in the preceding theorem.
If $\eta=\tau$, then all the nonzero real eigenvalues of $H$ are
negative. \item[] {\bf Proof:} For $\eta=\tau$, $\bbr
\psi_n,+|\psi_n,+\kkt_\eta>0$ whereas $\bbr
\psi_n,-|\psi_n,-\kkt_\eta<0$. Hence according to (\ref{spect}),
$E_n<0$.~~~$\square$
\end{itemize}

\section{Phermions and $N=2$ Pseudo-Supersymmetry with a Definite
Metric}

Consider realizing the algebra (\ref{N=2-ph-susy}) of $N=2$
pseudo-SQM using a Hamiltonian of the form~(\ref{H}) with $N$
being the boson number operator (\ref{boson-number}). Then in view
of the analogy with the boson-fermion oscillator and requiring
that the pseudo-supersymmetry is an exchange symmetry of a boson
and a particle having ${\cal N}$ as its number operator, we are
again led to an expression of the form~(\ref{q-susy}) for the
symmetry generator $\QQ$, namely
    \be
    \QQ=\sqrt{2E}~a^\dagger\alpha_+,~~~~~~~~E>0.
    \label{q-ph-susy}
    \ee
With this choice of $\QQ$, we may fulfill relations
(\ref{N=2-ph-susy}) provided that we require the undetermined
particle to be a pseudo-Hermitian fermion (phermion), whose
defining algebra and number operator are respectively given by
(\ref{normal-ph-fermion}) and (\ref{ph-fermion-number}), and that
we postulate the relative bose statistics, i.e., $a$ and
$a^\dagger$ commute with any operator constructed out of
$\alpha,\alpha^\dagger$, and $\eta$.

We can easily obtain a Fock-space representation of the phermion
algebra in the same way one obtains the Fock-space representation
of the fermion algebra \cite{fermion-algebra}. The representation
space ${\cal V}$ is the span of $|+\kt:=\alpha_+$ and
$|-\kt:={\cal N}_+=\alpha_+^\sharp\alpha_+$. This is a
two-dimensional subspace of the phermion algebra viewed as a
four-dimensional complex vector space (with
$\{1,\alpha_+,\alpha_+^\sharp,{\cal N}_+\}$ as a basis).
Therefore, ${\cal V}$ is isomorphic to $\C^2$ as a vector space.
In the basis $\{|\pm\kt\}$, the elements of ${\cal V}$ are
represented by column vectors, e.g.,
    \be
    |+\kt=\left(\begin{array}{c} 1 \\ 0 \end{array}\right),
    ~~~~~~~~~~
    |-\kt=\left(\begin{array}{c} 0 \\ 1 \end{array}\right),
    \label{states}
    \ee
and linear operators acting in ${\cal V}$ take the form of
$2\times 2$ matrices. In particular, if we denote the
representation map by $\rho_\star$, we find
    \be
    \rho_\star(\alpha_+)=\left(\begin{array}{cc}
            0 & 1 \\
            0 & 0\end{array}\right)=\alpha,~~~~
    \rho_\star(\alpha_+^\sharp)=\left(\begin{array}{cc}
            0 & 0 \\
            1 & 0\end{array}\right)=\alpha^\dagger,
    \label{f10}
    \ee
where we have made use of (\ref{normal-ph-fermion}) and
(\ref{f1}).

Equations (\ref{f10}) show that the phermion algebra
(\ref{normal-ph-fermion}) admits a faithful irreducible
representation $\rho_\star$ that is identical with the basic
representation $\rho_f$ of the fermion algebra (\ref{fermion-cr}).
This representation is a $*$-representation provided that $\C^2$
is equipped with the standard Euclidean inner product. This
happens if $\rho_\star(\eta)$ is just the identity matrix $I$
which in view of the fact that $\rho_\star$ is a faithful
representation implies $\eta=1$. But this corresponds to an
ordinary fermion.

There are also other choices for the inner product on $\C^2$ that
renders $\rho_\star$ a $*$-representation. Let $\rho_\star(\eta)$
be the associated metric operator. Then one can show using
(\ref{f10}) and
    \[\rho_\star(\eta)\rho_\star(\alpha_+^\sharp)=
    \rho_\star(\eta)\rho_\star(\eta^{-1}\alpha_+^\dagger\eta)=
    \rho_\star(\alpha_+)^\dagger\rho_\star(\eta)\]
that $\rho_\star(\eta)=s I$, where $s\in\R-\{0\}$. This in
particular shows that $\eta$ is proportional to $1$ and that in
effect a phermion is equivalent to an ordinary fermion.

The equivalence of a phermion and a fermion may be stated in terms
of the underlying abstract algebras: {\em The abstract phermion
algebra actually coincides with the abstract fermion algebra
${\cal F}$.} One way of seeing this is to note that any faithful
representation of the phermion algebra (\ref{normal-ph-fermion})
is completely reducible to copies of the above-described basic
representation $\rho_\star$.\footnote{This may be established
using the approach of \cite{jpa}.} Furthermore, the irreducible
$*$-representations of this algebra does not support an indefinite
metric operator. A more explicit demonstration of the latter
observation is provided by considering an arbitrary
two-dimensional faithful representation $\rho$ of
(\ref{normal-ph-fermion}), which is equivalent to $\rho_\star$,
and supposing that $\rho(\eta)$ is an indefinite operator, i.e.,
an indefinite invertible Hermitian matrix. One can then perform a
similarity transform $S:\C^2\to\C^2$ that transforms $\rho(\eta)$
as \cite{p4,p7}:
    \be
    \rho(\eta)\to \sigma(\eta):=S^\dagger\rho(\eta)S=
    \left(\begin{array}{cc}
            1 & 0 \\
            0 & -1\end{array}\right)=:\sigma_3.
    \label{s3}
    \ee
Under this transformation
    \be
    \rho(\alpha_+)\to \sigma(\alpha_+):=
    S^{-1}\rho(\alpha_+)S=\left(\begin{array}{cc}
            s & u \\
            v & t\end{array}\right),
    \label{s4}
    \ee
for some $s,t,u,v\in\C$. Now, in view of the identities
$\rho(\alpha_+)^2=0\neq\rho(\alpha_+)$, we have
$\sigma(\alpha_+)^2=0\neq \sigma(\alpha_+)$. The latter relations
fix $s$ and $t$ according to
    \be
    s=-t=\pm\sqrt{uv}.
    \label{s5}
    \ee
Substituting (\ref{s5}) in (\ref{s4}) and making use of
(\ref{s3}), we have
    \be
    \rho(\{\alpha_+,\alpha_+^\sharp\})=
    \{\rho(\alpha_+)\;,\;\rho(\eta)^{-1}
    \rho(\alpha_+)^\dagger\rho(\eta)\}=
    S\left\{\sigma(\alpha_+),\sigma_3\sigma(\alpha_+)^\dagger
    \sigma_3\right\}S^{-1}
    =-(|u|-|v|)^2I.
    \label{ab}
    \ee
In particular, $\rho(\{\alpha_+,\alpha_+^\sharp\})$ cannot be
equated to the identity matrix $I$ as required by
(\ref{normal-ph-fermion}). This shows that $\eta$ cannot be an
indefinite operator. That is the realization of $N=2$ pseudo-SQM
in terms of the boson-phermion oscillator only applies to the
cases that $\eta$ is a positive- or negative-definite operator.

For a negative-definite metric operator $\eta$, we can use the
positive-definite metric operator $-\eta$ to define the
pseudo-adjoint of the relevant operators. Hence without loss of
generality we will suppose that $\eta$ is positive-definite. In
this case $\eta$ has a unique positive-definite square root
$\eta^{1/2}$ with inverse $\eta^{-1/2}$. It is an easy exercise to
show that $\alpha_+$ and
$\alpha_+^\sharp=\eta^{-1}\alpha_+^\dagger\eta$ satisfy the
phermion algebra (\ref{normal-ph-fermion}) if and only if
$\alpha:=\eta^{1/2}\alpha_+\eta^{-1/2}$ and $\alpha^\dagger=
\eta^{1/2}\alpha_+^\sharp\eta^{-1/2}$ satisfy the fermion algebra
(\ref{fermion-cr}). Therefore, the phermion
(\ref{normal-ph-fermion}) and fermion (\ref{fermion-cr}) algebras
are related by a similarity transformation, and phermions have the
same physical properties as the ordinary fermions. Similarly, the
$N=2$ pseudo-SQM is just another representation of the ordinary
$N=2$ SQM.\footnote{This result is consistent with those of
Refs.~\cite{p2,jpa-2003} where the pseudo-Hermitian Hamiltonians
with a definite metric operator (that is the so-called
quasi-Hermitian Hamiltonians \cite{p7,cjp1}) are related to
Hermitian Hamiltonians via similarity transformations.}

\section{Abnormal Phermions and $N=2$ Pseudo-Supersymmetry with an
Indefinite Metric}

Because $N=2$ pseudo-supersymmetry algebra with an indefinite
metric cannot be realized using a boson-phermion oscillator, we
seek for a modification of the phermion algebra. A natural
modification is suggested by (\ref{ab}). It is the algebra of an
abnormal phermion (\ref{ph-fermion}).

Using the notion of an abstract creation and annihilation algebra
outlined in Section~2, we define an abnormal pseudo-Hermitian
fermion (abnormal phermion) as the abstract complex $*$-algebra
$\Phi$ generated by three elements: $c,c^*,n$ and subject to
relations (\ref{a1}) and
    \bea
    &&n=-\,c^*c,
    \label{fi-1}\\
    &&c^2=0,
    \label{fi-2}\\
    &&\{c,c^*\}=-1,
    \label{fi-3}
    \eea
together with a faithful irreducible Hilbert-space representation
$({\cal H},\rho)$ of $\Phi$ such that this representation is a
$*$-representation if we endow ${\cal H}$ with the (indefinite)
inner product $\bbr\cdot,\cdot\kkt_\eta$ for some indefinite
metric operator $\eta$ acting in ${\cal H}$.

Viewing $\Phi$ as a complex associative algebra, i.e.,
disregarding its $*$-structure, we may define
    \be
    c_1:=-ic,~~~~~~~~~c_2:=-ic^*,
    \label{complex}
    \ee
and check that $c_1,c_2$ and $n=c_2c_1$ satisfy the defining
relations of the fermion algebra ${\cal F}$, i.e., as complex
associate algebras $\Phi$ and ${\cal F}$ are isomorphic. This is
sufficient to conclude that $\Phi$ has a unique two-dimensional
faithful irreducible representation with representation map
$\varrho_\star:\Phi\to{\rm End}(\C^2)$ given by
    \be
    \varrho_\star(c)=\left(\begin{array}{cc}
            0 & i \\
            0 & 0\end{array}\right)=i\alpha=:\alpha_-,~~~~~~~~
    \varrho_\star(c^*)=\left(\begin{array}{cc}
            0 & 0 \\
            i & 0\end{array}\right)=i\alpha^\dagger=
            -\alpha_-^\dagger.
    \label{fi-rep-1}
    \ee
Clearly, this is not a $*$-representation if we identify the inner
product on $\C^2$ with the Euclidean inner product. But it is a
$*$-representation if we endow $\C^2$ with the indefinite inner
product $\bbr\cdot,\cdot\kkt_{\sigma_3}$ where $\sigma_3$ is the
diagonal Pauli matrix (\ref{s3}). Using this inner product to
define the pseudo-adjoint $\sharp$, we have
$\varrho_\star(c)^\sharp=
\sigma_3\varrho_\star(c)^\dagger\sigma_3=\varrho_\star(c^*)$ or
simply
    \be
    \alpha_-^\sharp=\varrho_\star(c^*).
    \label{fi-rep-2}
    \ee
With $\alpha_-$ and $\alpha_-^\sharp$ given by (\ref{fi-rep-1})
and (\ref{fi-rep-2}) and recalling that they provide the unique
faithful irreducible $*$-representation of $\Phi$ we will
respectively identify $c$ and $c^*$ with $\alpha_-$ and
$\alpha_-^\sharp$, speak of (\ref{ph-fermion}) as the abnormal
phermion algebra, and let
    \be
    {\cal N}_-:=-\alpha_-^\sharp\alpha_-
    \label{ab-ph-number}
    \ee
be the abnormal phermion number operator.

Now, we are in a position to explore the Hamiltonian~(\ref{H})
with $N$ and ${\cal N}$ respectively identified with the boson
number operator (\ref{boson-number}) and the abnormal phermion
number operator (\ref{ab-ph-number}). Postulating the relative
bose statistics, i.e., that $a$ and $a^\dagger$ commute with any
operator constructed out of $\alpha_-$, $\alpha_-^\sharp$, and
$\eta=\sigma_3$, we can easily check that $H$ together with
    \be
    \QQ=:\sqrt{2|E|}~a^\dagger\alpha_-~~~~~~~{\rm and}~~~~~E<0
    \label{q-ab-ph-susy}
    \ee
satisfy the algebra (\ref{N=2-ph-susy}) of $N=2$ pseudo-SQM with
the indefinite metric operator $\eta=\sigma_3$. Note that the
grading operator for this system is again given by $\tau=1-2{\cal
N}_-=\sigma_3$. Therefore, the hypothesis ($\eta=\tau$) of the
Corollary given in Section~3 is satisfied and the real eigenvalues
of $H$ must be negative. Indeed it is not difficult to check that
the eigenvalues of $H$ are real and non-positive. They are given
by  $E_n=nE=-n|E|$ where $n\in\N$.\footnote{The fact that energy
spectrum and therefore the Hamiltonian is bounded above but not
below may be used to argue that this Hamiltonian does not describe
a physical system. The problem with arbitrarily large negative
energies may be avoided using Feynman's idea of associating this
Hamiltonian with a system that evolves backward in time. This is
equivalent to considering $-H$ as the Hamiltonian for the
corresponding forward evolution in time. Although $-H$ coincides
with the boson-fermion Hamiltonian, it describes a fundamentally
different system as we explain below.}

The realization of the $N=2$ pseudo-SQM in terms of a
boson-abnormal-phermion exchange symmetry as outlined above enjoys
a uniqueness property in the sense that considering an arbitrary
two-dimensional irreducible representation of $\Phi$ with
arbitrary indefinite metric operator ($2\times 2$ indefinite
invertible Hermitian matrix) on the representation space leads to
an equivalent description of the abnormal phermion and the
associated boson-abnormal-phermion exchange pseudo-supersymmetry.
This is because, as noted in Section~4, any such metric operator
may be transformed to $\sigma_3$ via a similarity transformation
of the Hilbert space \cite{p4,p7}.

It is not difficult to see that if we adopt the basic
representations of ordinary fermions $\rho_\star$ and abnormal
phermions $\varrho_\star$ with the metric operators given by
$\rho_\star(\eta)=I$ and $\varrho_\star(\eta)=\sigma_3$, then the
corresponding number operators coincide: ${\cal N}_-={\cal N}_+$.
This does not however mean that a fermion and an abnormal phermion
are equivalent. The distinction lies in the interpretation of the
abnormal phermion state described by the state vector $|-\kt$ of
(\ref{states}) that satisfy $\bbr-,-\kkt_{\sigma_3}=
\br-|\sigma_3|-\kt=-1$. This state vector does not belong to the
physical Hilbert space, for the latter includes besides the zero
vector only the state vectors with positive real norms
\cite{sudarshan-61,ph-indefinite,cjp1}. Therefore, unlike a
fermion that has two physical states, an abnormal phermion has a
single physical state. This in turn implies that an ordinary
quantum mechanical system consisting of only a single abnormal
phermionic degree of freedom is trivial. Nontrivial systems may
however be constructed by combining an abnormal phermion with
other particles or using more than one abnormal
phermion.\footnote{Alternatively, one may associate the unphysical
state with the physical state of another particle, e.g., the
corresponding anti-particle.}

As an example consider a system consisting of $\ell$ abnormal
phermions with annihilation, creation, number, and metric
operators $\alpha_-^{(i)}$, $\alpha_-^{(i)\sharp}$, ${\cal
N}_-^{(i)}:=-\alpha_-^{(i)\sharp} \alpha_-^{(i)}$, and
$\eta^{(i)}$, respectively. Suppose that for all
$i\in\{1,2,\cdots,\ell\}$, $\eta^{(i)}=\sigma_3$. Clearly, the
Hilbert space of this system is $2^\ell$ dimensional; a set of
basis vectors are given by:
    \be
    |\nu_1,\nu_2,\cdots,\nu_\ell\kt:=
    (\alpha_-^{(1)\sharp})^{\nu_1}
    (\alpha_-^{(2)\sharp})^{\nu_2}
    \cdots (\alpha_-^{(\ell)\sharp})^{\nu_\ell}
    |0,0,\cdots,0\kt,
    \label{many}
    \ee
where $\nu_i\in\{0,1\}$, $|0,0,\cdots,0\kt:=|0\kt^{(1)}\otimes
|0\kt^{(2)}\otimes\cdots\otimes |0\kt^{(\ell)}$, is the vacuum
state vector for the system, $|0\kt^{(i)}$ is the vacuum state
vector for the $i$-th abnormal phermion:
$\alpha_-^{(i)}|0\kt^{(i)}=0$, and we adopt the relative fermi
statistics, i.e., for all for $i,j\in\{1,2,\cdots,\ell\}$,
    \be
    \{\alpha_-^{(i)},\alpha_-^{(j)}\}=0,~~~~~~~~~~
    \{\alpha_-^{(i)},\alpha_-^{(j)\sharp}\}=-\delta_{ij}.
    \label{rel-fermi}
    \ee

The $\eta$-inner product of two basis vectors (\ref{many}) is
defined according to
    \bea
    \bbr\mu_1,\mu_2,\cdots,\mu_\ell|
    \nu_1,\nu_2,\cdots,\nu_\ell\kkt_\eta&:=&
    \br\mu_1|\eta^{(1)}|\nu_1\kt\br\mu_2|\eta^{(2)}|\nu_2\kt
    \cdots \br\mu_\ell|\eta^{(\ell)}|\nu_\ell\kt\nn\\
    &=&(-1)^{\nu_1+\nu_2+\cdots+\nu_\ell}\delta_{\mu_1,\nu_1}
    \delta_{\mu_2,\nu_2}\cdots\delta_{\mu_\ell,\nu_\ell},
    \label{inner}
    \eea
where $|\nu_i\kt:=(\alpha_-^{(1)\sharp})^{\nu_i}|0\kt^{(i)}$. In
particular, the state vectors associated with an even number of
particles have a positive real $\eta$-norm. These span the
physical Hilbert space ${\cal H}_{\rm phys}$ of the system which
is $2^{\ell-1}$-dimensional. The physical state vectors may be
constructed from the vacuum state vector $|0,0,\cdots,0\kt$ and
the `physical' creation operators: $\alpha_{ij}^+:=
\alpha_-^{(j)\sharp}\alpha_-^{(i)\sharp}$ with $i<j$. These
together with the `physical' annihilation operators $\alpha_{ij}:=
\alpha_-^{(i)}\alpha_-^{(j)}$ satisfy, for all $i<j$ and $k<l$,
    \bea
    [\alpha_{ij},\alpha_{kl}]&=&[\alpha_{ij}^+,\alpha_{kl}^+]=0,
    \label{phys-1}\\
    \left[\alpha_{ij},\alpha_{kl}^+\right]&=&\delta_{ik}\delta_{jl}-
    \delta_{ij}\delta_{jk}+
    \delta_{ik}\beta_{lj}+
    \delta_{jl}\beta_{ki}-
    \delta_{jk}\beta_{li}-
    \delta_{il}\beta_{kj},
    \label{phys-2}
    \eea
where we have made use of (\ref{rel-fermi}) and introduced
    \[ \beta_{ij}:=\alpha_-^{(i)\sharp}\alpha_-^{(j)}.\]
Note that the shift operators $\beta_{ij}$ commute with the total
number operator ${\cal N}_{\rm tot}=\sum_{i=1}^\ell {\cal
N}_-^{(i)}$. Hence they relate different (physical) states with
the same number of particles.

Clearly, the same physical Hilbert space may be obtained using a
system of $\ell-1$ fermions. But then different states are related
by fermionic creation and annihilation operators that satisfy
anti-commutation relation. In contrast, the above description
using the abnormal phermions leads to a set of creation,
annihilation, and shift operators that satisfy commutation
relations. As a result, it makes the underlying (Lie) group
structure and the associated symmetries of the system more
transparent. The use of abnormal phermions seems to provide a
certain type of `bosonization' of the fermionic
systems.\footnote{A similar argument applies to a parafermionic
description of the physical Hilbert space using a parafermion of
order $2^{\ell-1}-1$. But the corresponding operator algebra would
involve complicated ternary relations. One may also try to obtain
a realization of parafermionic operators (with the above order) in
terms of the abnormal phermionic operators. This may be viewed as
a `bosonization' of the associated parafermionic system.}

\section{Group Theoretical Basis of Phermion and Abnormal
Phermion Algebras}

The algebras of the phermions (\ref{normal-ph-fermion}) and
abnormal phermions (\ref{ph-fermion}) may be expressed in the
unified form
    \be
    \alpha_\epsilon^2=0,~~~~~~~~
    \{\alpha_\epsilon,\alpha_\epsilon^\sharp\}=\epsilon\,1,
    \label{unif-1}
    \ee
with $\epsilon\in\{-,+\}$. The phermion (\ref{ph-fermion-number})
and abnormal phermion (\ref{ab-ph-number}) number operators,
    \be
    {\cal N}_\epsilon=\epsilon\;\alpha_\epsilon^\sharp\alpha_\epsilon,
    \label{unif-2}
    \ee
satisfy
    \be
    [\alpha_\epsilon,N_\epsilon]=\alpha_\epsilon,~~~~~~~
    [\alpha_\epsilon^\sharp,N_\epsilon]=-\alpha_\epsilon^\sharp.
    \label{unif-3}
    \ee
Furthermore, in the basic two-dimensional representations
$\rho_\star$ and $\varrho_\star$ with $\rho_\star(\eta)=1$ for
phermion (so that it is just a fermion) and
$\varrho_\star(\eta)=\sigma_3$ for the abnormal phermion, we can
easily check using (\ref{f1}), (\ref{fi-rep-1}) and
(\ref{fi-rep-2}) that
    \be
    [\alpha_\epsilon,\alpha_\epsilon^\sharp]=
    1-2\epsilon {\cal N}_\epsilon.
    \label{unif-4}
    \ee

Now, let us introduce the pseudo-Hermitian operators \cite{p1}
    \be
    J^\epsilon_1:=
    \frac{1}{2}\,(\alpha_\epsilon+\alpha_\epsilon^\sharp),~~~~~~
    J^\epsilon_2:=
    \frac{1}{2i}\,(\alpha_\epsilon-\alpha_\epsilon^\sharp),~~~~~~
    J^\epsilon_3:=-{\cal N}_\epsilon+\frac{1}{2},
    \label{SU}
    \ee
and express (\ref{unif-3}) and (\ref{unif-4}) in terms of $J_{\pm
i}$. This yields, for all $i,j\in\{1,2,3\}$,
    \be
    [J^\epsilon_i,J^\epsilon_j]=i
    \sum_{k=1}^3 \delta^\epsilon_{k}\,\epsilon_{ijk}J^\epsilon_k,
    \label{su-2,11}
    \ee
where $\delta_{k}^+:=1$, $\delta_{k}^-:=(1-2\delta_{3,k})$,
$\delta_{i,j}$ is the Kronecker delta function, and
$\epsilon_{ijk}$ is the totally antisymmetric Levi-Civita symbol
(with $\epsilon_{123}=1$).

Equations (\ref{su-2,11}) are the defining relations for the Lie
algebra $su(2)=so(3)$ for $\epsilon=+$ (i.e., for
phermion/fermion) and $su(1,1)=so(2,1)$ for $\epsilon=-$ (i.e.,
for abnormal phermion). This yields a direct correspondence
between the abstract fermion algebra ${\cal F}$ and the Lie
algebra $su(2)$ and the abstract abnormal-phermion algebra $\Phi$
and the Lie algebra $su(1,1)$. It further suggests a potential
application of abnormal phermions in the study of
Klein-Gordon-type field equations \cite{cqg,ap}, for the effective
Hamiltonian in the two-component formulation of these equations
involves the elements of $su(1,1)$, \cite{jmp-98}.

\section{Conclusion}

The recent study of pseudo-Hermitian operators \cite{p1,p2} has
its root in an attempt to understand the mathematical origin of
the surprising spectral properties of $PT$-symmetric Hamiltonians
\cite{bender}. It has not only provided means for a more realistic
assessment of the role of $PT$-symmetry (and other antilinear
symmetries) \cite{p3,jpa-2003} but has found applications in
various other problems \cite{cqg,ap,applications}. Among the most
natural outcomes of this study is the formulation of the
pseudo-supersymmetric quantum mechanics \cite{p1,p4}. This is a
genuine generalization of ordinary supersymmetric quantum
mechanics with similar topological properties \cite{p4}. It allows
for a more general class of factorizations of a given
pseudo-Hermitian and in particular Hermitian Hamiltonian. In this
article we elucidated the consequences of the presence of negative
real eigenvalues for a pseudo-supersymmetric Hamiltonian and
explored the statistical origin of the pseudo-supersymmetric
quantum mechanics.

The simplest oscillator realizations of pseudo-supersymmetry
involve the phermionic and abnormal phermionic degrees of freedom
depending on whether the associated metric operator is definite or
indefinite. We showed that a phermion is physically equivalent to
an ordinary fermion. The situation is quite different for an
abnormal phermion, for only half of the states of abnormal
phermionic systems correspond to physical states. The latter
correspond to states with an even number of abnormal phermions.
They are related via a set of composite creation and annihilation
operators that together with a set of shift operators satisfy
certain commutation relations. The physical sector of a system of
$\ell$ abnormal phermions may be described by $\ell-1$ ordinary
fermions. However the latter description involves anticommutation
relations and makes the study of the underlying Lie group
structure of the system obscure.

Another interesting outcome of the present study is related to the
association of the compact Lie algebra $su(2)=so(3)$ with ordinary
fermions and the noncompact Lie algebra $su(1,1)=so(2,1)$ with
abnormal phermions. This suggests a possible application of
abnormal phermions in physical problems that have $su(1,1)$ as a
kinematical, dynamical, or symmetry group \cite{nova}.

The introduction of the concept of an abnormal phermion as offered
in the present paper raises various related issues. We close this
paper by commenting on a few of the most notable of these.
    \begin{enumerate}
    \item {\bf Classical abnormal phermionic degrees of freedom
and their quantization:} Similarly to the case of abnormal bosonic
degrees of freedom \cite{sudarshan-61}, abnormal phermions and
normal phermions (fermions) both have the usual Grassmann (odd
super number \cite{supermanifolds}) variables as their classical
counterpart. The choice of abnormal anticommutation
relations~(\ref{ph-fermion}) over the normal anticommutation
relations~(\ref{fermion-cr}) may be viewed as an alternative way
of quantizing a classical fermionic degree of freedom. In
particular, one may quantize a classical supersymmetric system to
obtain a quantum pseudo-supersymmetric system by employing
abnormal quantization scheme for classical fermionic degree(s) of
freedom and normal quantization scheme for the bosonic degree(s)
of freedom. Indeed, an interesting direction of further research
would be to use the approach of Ref.~\cite{sudarshan-61} to
construct concrete examples of systems with normal and abnormal
phermionic degrees of freedom and investigate their field
theoretical analogs.
    \item {\bf Abnormal phermion-abnormal boson exchange symmetry:}
By conducting abnormal quantization of both fermionic and bosonic
degrees of freedom of a classical supersymmetric system one
obtains a quantum system with abnormal phermionic and abnormal
bosonic degrees of freedom and a quantum analog of the classical
supersymmetry transformations relating them. A natural question is
to investigate the nature (operator algebra) of this symmetry.
    \item {\bf Multi abnormal phermionic versus parafermionic systems:}
One can use the physical creation and annihilation operators for
multi abnormal phermionic systems to obtain a realization of those
of a parafermionic system of appropriate order. This may be viewed
as an alternative to the Green's ansatz
\cite{green,greenberger-messiah} and orthofermionic
\cite{jpa,top-susy3} constructions of the latter. It may also
provide means to describe the hidden supersymmetries of the
corresponding parafermionic systems \cite{hidden} and investigate
their analogs for general multi abnormal phermionic systems.
    \end{enumerate}
\vspace{.3cm}

{\bf Note:} After the completion of this project
Ref.~\cite{habara} was brought to my attention (by one of the
referees) where the authors use the abnormal bosonic degrees of
freedom to formulate a bosonic analog of the Dirac sea for
fermions. The complexification scheme used in \cite{habara} is
similar to the one given in Eqs.~(\ref{complex}).

\section*{Acknowledgment}

This work has been supported by the Turkish Academy of Sciences in
the framework of the Young Researcher Award Program
(EA-T$\ddot{\rm U}$BA-GEB$\dot{\rm I}$P/2001-1-1).

\ed
{\small
\begin{thebibliography}{99}
\bibitem{susy-qft} J.~Wess and B.~Zumino, Nucl.~Phys.~B {\bf 70}, 39
(1974).
\bibitem{witten-81-82} E.~Witten, Nucl.~Phys.~B {\bf 185}, 513
(1981); ibid {\bf 202}, 253 (1982).
\bibitem{susy-review} L.~E.~Gendenshtein and I.~V.~Krive, Sov.\
Phys.\ Usp.~{\bf 28}, 645-666 (1985).
\bibitem{susy-review2} F.~Cooper, A.~Khare, and U.~Sukhatme,
Phys.\ Rep.\ {\bf 251}, 267-385 (1995).
\bibitem{susy-spin} H.~Nicolai, J.~Phys.~A: Math.\ Gen.~{\bf 9},
1497 (1976).
\bibitem{susy-book} G.~Junker, {\em Supersymmetric Methods in
Quantum and Statistical Physics} (Springer-Verlag, Berlin, 1996).
\bibitem{susy-math} E.~Witten, J.~Diff.~Geometry {\bf 17}, 661
(1982);\\
L.~Alvarez-Gaume, Commun.\ Math.\ Phys.\ {\bf 90}, 161 (1983);\\
L.~Alvarez-Gaume, J.\ Phys.\ A: Math.\ Gen.\ {\bf 16}, 4177
(1983);\\
P.~Windey, Acta.\ Phys.\ Pol.\ B {\bf 15}, 453 (1984);\\
A.~Mostafazadeh, J.\ Math.\ Phys.\ {\bf 35}, 1095 (1994).
\bibitem{para-susy} V.~A.~Rubakov and V.~P.~Spiridonov, Mod.\ Phys.\ Lett.\
{\bf A 3}, 1337 (1988);\\
J.~Beckers and N.~Deberg, Nucl.\ Phys.\ B {\bf 340}, 767 (1990);\\
S.~Durand, M.~Mayrand and V.~Spridonov, Mod.\ Phys.\ Lett.\ A {\bf
6}, 3163 (1991);\\
M.~Tomiya, J.~Phys.\ A: Math.\ Gen.\ {\bf 25}, 4699 (1992);\\
A.~Khare, J.~Phys.\ A: Math.\ Gen.\ {\bf 25}, L749 (1992);\\
A.~Khare, J.\ Math.\ Phys. {\bf 34}, 1277 (1993);\\
A.~Mostafazadeh, Int.~J.~Mod.\ Phys.~A {\bf 11}, 1057 (1996).
\bibitem{ortho-susy} A.~Khare, A.~K.~Mishra and G.~Rajasekaran,
Int.\ J.\ Mod.\ Phys.\ A {\bf 8},1245 (1993).
\bibitem{jpa} A.~Mostafazadeh, J.~Phys.~A: Math.\ Gen.\ {\bf 34},
8601 (2001).
\bibitem{q-susy} V.~Spiridonov,  Mod.\ Phys.\ Lett.\ A {\bf 7},
1241 (1992);\\
W.~-S.~Chung, Phys.~Lett.~A {\bf 259}, 437 (1999).
\bibitem{f-susy} C.~Ahn, D.~Bernard, and A.~Leclair, Nucl.\ Phys.~B
{\bf 346}, 409 (1990);\\
L.~Baulieu and E.~G.~Floratos, Phys.\ Lett.~B {\bf 258}, 171
(1991);\\
R.~Kerner, J.~Math.\ Phys.\ {\bf 33}, 403 (1992);\\
S.~Durand, Phys.\ Lett.\ B {\bf 312}, 115 (1993);\\
S.~Durand, Mod.\ Phys.\ Lett.\ A {\bf 8}, 1795 (1993);
ibid 2323 (1993);\\
A.~T.~Filippov, A.~P.~Isaev, and R.~D.~Kurdikov, Mod.\ Phys.\
Lett.~A {\bf 7}, 2129 (1993);\\
N.~Mohammedi,  Mod.\ Phys.\ Lett.~A {\bf 10}, 1287 (1995);\\
N.~Fleury and M.~Rausch~de~Traubenberg, Mod.\ Phys.\ Lett.~A {\bf
11}, 899 (1996).
\bibitem{ijmpa-96-3} A.~Mostafazadeh, Int.~J.~Mod.\ Phys.\ Lett~A
{\bf 17}, 2957 (1996).
\bibitem{qb-f} W.~-S.~Chung, Prog.\ Theor.\ Phys.~{\bf 94}, 649
(1995).
\bibitem{qb-qf} R.~Parthasarathy and K.~S.~Viswanathan, J.~Phys.~A:
Math.\ Gen.\ {\bf 24}, 613 (1991);\\
W.~-S.~Chung, J.~Phys.\ A: Math.\ Gen.\ {\bf 32}, 2605 (1999).
\bibitem{top-susy1} A.~Mostafazadeh and K.~Aghababaei Samani,
Mod.\ Phys.\ Lett.\ A {\bf 15}, 175 (2000).
\bibitem{top-susy2} K.~Aghababaei Samani and A.~Mostafazadeh ,
Nucl.\ Phys.~B {\bf 595},467 (2001).
\bibitem{top-susy3} K.~Aghababei~Samani and A.~Mostafazadeh,
Mod.\ Phys.\ Lett.~A {\bf 17}, 131 (2002).
\bibitem{top-susy4} A.~Mostafazadeh, Nucl.~Phys.~B {\bf 624}, 500
(2002);\\
C.~Quesne, Mod.\ Phys.\ Lett.~A {\bf 18}, 515 (2003).
\bibitem{p1} A.\ Mostafazadeh, J.\ Math.\ Phys.\ {\bf 43}, 205
(2002).
\bibitem{p4} A.~Mostafazadeh, Nucl.\ Phys.\ B {\bf 640},
419 (2002).
\bibitem{old} F.~Cannata, G.~Junker, J.~Trost, Phys.\ Lett.~A
{\bf 246}, 219 (1998);\\
A.~Andrianov, M.~V.~Ioffa, F.~Cannata, J.-P.~Dedonder,
Int.~J.~Mod.\ Phys.~A {\bf 14}, 2675 (1999);\\
M.~Znojil, F.~Cannata, B.~Baghci, and R.~Roychoudhury, Phys.\
Lett.~B {\bf 483}, 284 (2000).
\bibitem{indefinite} J.\ Bogn\'ar, {\em Indefinite Inner Product
Spaces} (Springer, Berlin, 1974);\\
T.\ Ya.\ Azizov and I.\ S.\ Iokhvidov, {\em Linear Operators in
Spaces with Indefinite Metric} (Wiley, Chichester, 1989).
\bibitem{beckers} J.~Beckers and N.~Deberg, Int.~J.~Mod.\ Phys.\
Lett~A {\bf 10}, 2783 (1995).
\bibitem{qv} C.~Quesne and N.~Vansteenkiste,
Mod.\ Phys.\ Lett.\ A {\bf 18}, 271 (2003).
\bibitem{sudarshan-61} E.~C.~G.~Sudarshan, Phys.\ Rev.\ {\bf 123}, 2183
(1961).
\bibitem{ph-indefinite} W.\ Pauli, Rev.\  Mod.\ Phys., {\bf 15}, 175
(1943);\\
T.~D.~Lee and G.~C.~Wick, Nucl.\ Phys~B {\bf 9}, 209 (1969).
\bibitem{cjp1} A.~Mostafazadeh, Czech J.~Phys.~{\bf 53}, 1079
(2003).
\bibitem{fell-doran} J.~M.~G.~Fell and R.~S.~Doran,
{\em Representations of $*$-Algebras, Locally Compact Groups, and
Banach $*$-Algebraic Bundles} (Academic Press, San Diego, 1988).
\bibitem{para} H.~S.~Green, Phys.\ Rev.\ {\bf 90}, 270 (1953).
\bibitem{greenberger-messiah} O.~W.~Greenberger and
A.~M.~L.Messiah, Phys.\ Rev.\ {\bf 138}, 1155 (1965).
\bibitem{q-deformed-boson}
M.~Arik and D.~D.~Coon, J.~Math.~Phys.\ {\bf 17}, 524 (1976);\\
A.~J.~Macfarlane, J.~Phys.~A: Math.\ Gen.\ {\bf 22}, 4581
(1989);\\
L.~C.~Biedenharn, J.~Phys.~A: Math.\ Gen.\ {\bf 22}, L873
(1989);\\
C.-P.~Sun and H.-C.~Fu, J.~Phys.~A: Math.\ Gen.\ {\bf 22}, L983
(1989).
\bibitem{general} A.~K.~Mishra and G.~Rajasekaran, Pramana
J.~Phys.\ {\bf 40}, 149 (1993).
\bibitem{supermanifolds} B.~DeWitt, {\em Supermanifolds},
2nd Ed.\ (Cambridge Uni.\ Press, Cambridge, 1992).
\bibitem{non-lin-susy} A.~A.~Andrianov, M.~V.~Ioffe, and
V.~P.~Spiridonov, Phys.\ Lett.~A {\bf 174}, 273 (1993).
\bibitem{fermion-algebra} A.~Mostafazadeh, preprint:
math-ph/0312065.
\bibitem{p7} A.~Mostafazadeh, J.\ Math.\ Phys.\ {\bf 44}, 974 (2003).
\bibitem{p2} A.~Mostafazadeh, J.\ Math.\ Phys.\ {\bf 43}, 2814 (2002).
\bibitem{jpa-2003} A.~Mostafazadeh, J.~Phys.~A: Math.\ Gen.\
{\bf 36}, 7081 (2003).
\bibitem{cqg} A.~Mostafazadeh, Class.\ Quantum Grav.\ {\bf 20}, 155
(2003).
\bibitem{ap} A.~Mostafazadeh, Ann.~Phys.~(N.Y.) {\bf 309}, 1
(2004).
%\bibitem{foldy} K.~M.~Case, Phys.\ Rev.\ {\bf 95}, 1323
%(1954);\\
%L.~L.~Foldy, Phys.\ Rev.\ {\bf 102}, 568 (1956);\\
%H.\ Feshbach and F.\ Villars, Rev.\  Mod.\ Phys.\ {\bf 30}, 24
%(1958).
\bibitem{jmp-98} A. Mostafazadeh, J.\ Math.\ Phys.\ {\bf 39},
4499 (1998).
\bibitem{bender} C.~M.~Bender and S.~Boettcher, Phys.\ Rev.\ Lett.\
{\bf 80}, 5243 (1998);\\
C.~M.~Bender, S.~Boettcher, and P.~N.~Meisenger, J.~Math.\ Phys.\
{\bf 40}, 2201 (1999).
\bibitem{p3} A.\ Mostafazadeh, J.\ Math.\ Phys., {\bf 43}, 3944
(2002).
\bibitem{applications}
S.~M.~Klishevich and M.~Plyushchay, Nucl.\ Phys.~B {\bf 628}, 217
(2002);\\
R.~N.~Deb, A.~Khare, and B.~D.~Roy, Phys.\ Lett.~A {\bf 307}, 215
(2003);\\
F.~Cannata, M.~V.~Ioffe and D.~N.~Nishnianidze, Phys.\ Lett.~A
{\bf 310}, 344 (2003);\\
Z.\ Ahmed and S.\ R.\ Jain, Phys.\ Rev.~E {\bf 67}, 045106 (2003);
and J.~Phys.~A: Math.\ Gen {\bf 36}, 9711 (2003);\\
Z.\ Ahmed, Phys.\ Lett.~A {\bf 310}, 139 (2003).
\bibitem{nova} A.~Mostafazadeh, {\em Dynamical Invariants,
Adiabatic Approximation, and the Geometric Phase} (Nova Science
Publishers, New York, 2001).
\bibitem{green} H.~S.~Green, Phys.\ Rev.~{\bf 90}, 270 (1953).
\bibitem{hidden} S.~Klishevich and M.~Plyushchay,
Mod.~Phys.~Lett.~A {\bf 14}, 2739 (1999).
\bibitem{habara} Y.~Habara, H.~B.~Nielsen, and M.~Ninomiya,
preprint: hep-th/0312302.
\end{thebibliography}
}
